\documentclass[a4paper,11pt]{article}
\usepackage{pos}

\title{Quarkonium transport in strongly coupled plasmas}
\author[a]{Govert Nijs}
\author*[a]{Bruno Scheihing-Hitschfeld}
\author[b]{Xiaojun Yao}

\affiliation[a]{Center for Theoretical Physics, Massachusetts Institute of Technology,\\
  Cambridge, MA  02139, USA}  

\affiliation[b]{InQubator for Quantum Simulation, Department of Physics, University of Washington,\\
Seattle, WA 98195, USA}

\emailAdd{govert@mit.edu}
\emailAdd{bscheihi@mit.edu}
\emailAdd{xjyao@uw.edu}

\abstract{
Suppression of open heavy flavors and quarkonia in heavy-ion collisions is among the most informative probes of the quark-gluon plasma. Interpreting the full wealth of data obtained from the collision events requires a precise understanding of the evolution of heavy quarks and quarkonia as they propagate through the nearly thermal and strongly coupled plasma. Systematic theoretical studies of quarkonium time evolution in the QGP in the regime where the temperature of the QGP is much smaller than the inverse of quarkonium size have only been carried out in the past few years.
Such calculations require the evaluation of a gauge-invariant correlator of chromoelectric fields dressed with Wilson lines, which is similar to, but different from, the correlation used to define the well-known~\cite{Casalderrey-Solana:2006fio} heavy quark diffusion coefficient. The origin of this difference has been explained in~\cite{Binder:2021otw,Scheihing-Hitschfeld:2022xqx,Scheihing-Hitschfeld:2023tuz}. Here we show the results of the calculation of the analogous correlator in strongly coupled $\mathcal{N}=4$ SYM using the AdS/CFT correspondence at a finite temperature~\cite{Nijs:2023dks}. While it resembles the open heavy quark case, it has some crucial differences that highlight the relevance of quantum color correlations. We will also discuss the results for the quarkonium transport coefficients obtained from this correlator, thereby establishing the first analytic results at strong coupling in this context. We find that they vanish in the strong coupling limit for an $\mathcal{N}=4$ SYM plasma with a large number of colors.}

\FullConference{% HardProbes2023\\
 26-31 March 2023 \\
 Aschaffenburg, Germany\\
 ~\\
MIT-CTP/5584, IQuS@UW-21-057\\}

\begin{document}
\maketitle

\section{Quarkonium transport at high and low temperatures}

Bound systems of a heavy quark-antiquark ($Q\bar{Q}$) pair, i.e., quarkonia, constitute a dynamical probe of the quark-gluon plasma (QGP) at temperatures reached in heavy-ion collisions. Nonrelativistically, quarkonium is characterized by the hierarchy of scales $M\gg \frac{1}{r}\gg |E_b|$\@. These scales are the heavy quark mass $M$, the inverse of quarkonium size $\frac{1}{r}$ and the binding energy $|E_b|$\@. Depending on where the plasma temperature $T$ fits into this hierarchy, quarkonium may probe different features of the QGP.
In current heavy ion collision experiments, the highest temperature achieved is on the order of $500$\,MeV, which is smaller than the charm quark mass $M_c\approx1.3$\,GeV and much smaller than the bottom quark mass $M_b\approx4.2$\,GeV. Therefore, it is necessary to understand quarkonia in the regime $M\gg T$. However, the temperature may be larger or smaller than $1/r$, $|E_b|$\@.

When the QGP temperature is high $M\gg T \gg \frac{1}{r} \gg |E_b|$, the interaction between a $Q\bar{Q}$ pair is significantly screened. Therefore, such a $Q\bar{Q}$ pair can be described in terms of two independent heavy quarks that diffuse in the plasma. This dynamics can be approximately described by a Langevin equation with drag and diffusion, characterized by the heavy quark diffusion coefficient~\cite{Casalderrey-Solana:2006fio}.

On the timeline of a heavy ion collision, the temperature eventually drops to $T\sim\frac{1}{r}$ (unless its initial value is even lower). Consequently, the interaction between a $Q\bar{Q}$ pair can no longer be neglected and must be included in the Langevin description, which generates some correlation between the pair. When the temperature drops further, $M\gg  \frac{1}{r} \gg T \gg |E_b|$, color correlations between the pair become important, and their dynamics can be described by a Lindblad equation~\cite{Brambilla:2016wgg-2017zei}, which involves two real transport coefficients $\kappa_{\rm adj}$ and $\gamma_{\rm adj}$. They are defined in terms of a chromoelectric field correlator
\begin{align}
\label{eqn:kappa_gamma_adj}
\kappa_{\rm adj} + i\gamma_{\rm adj} = \frac{g^2 T_F }{3 N_c} \int d t\, \big\langle \hat{\mathcal{T}} E^a_i(t) \mathcal{W}^{ab}(t,0) E^b_i(0) \big\rangle_T \, ,
\end{align}
where $T_F$ is defined by ${\rm Tr}(T^a_FT^b_F)=T_F\delta^{ab}$. $T_F^a$ is the generator in the fundamental representation and $\mathcal{W}^{ab}(t,0)$ denotes a timelike Wilson line in the adjoint representation from time $0$ to $t$. For our purposes, we consider Wilson lines given by $\mathcal{W}(x,y) = {P} \exp \left( ig \int_y^x \!\! d z^\mu A_\mu^a(z) T^a_A \right)$, in which $x$ and $y$ are position 4-vectors connected by a straight line path. The expectation value of an operator $O$ is defined as $\langle O \rangle_T \equiv {\rm Tr}(O e^{-\beta H_E})/{\rm Tr}(e^{-\beta H_E})$, where $\beta = 1/T$ is the inverse of the plasma temperature and $H_E$ denotes the Hamiltonian of light quarks and gluons in the QGP\@.

As the QGP temperature continues dropping and finally becomes of the same order as the binding energy, $M\gg  \frac{1}{r} \gg T \sim |E_b|$, a semiclassical description in terms of a Wigner distribution becomes applicable. If we integrate said Wigner distribution over momentum, we arrive at a rate equation for the number density $n_b$ of a quarkonium state with quantum numbers $b$~\cite{Yao:2018nmy},
\begin{align}
\label{eqn:rate}
\frac{d n_b(t,{\bf x})}{d t} = -\Gamma\, n_b(t,{\bf x}) + F(t,{\bf x}) \,,
\end{align}
where $\Gamma$ is the dissociation rate and $F$ denotes the contribution of quarkonium formation (in-medium recombination). They are given by
\begin{align}
\Gamma &=  \frac{g^2T_F}{3N_c}\int \frac{d^3p_{{\rm rel}}}{(2\pi)^3} 
| \langle \psi_b | {\bf r} | \Psi_{{\bf p}_{\rm rel}} \rangle |^2 [g^{++}_E]^{>}\Big(-|E_b| - \frac{p^2_{\rm rel}}{M}\Big) \\
F(t,{\bf x}) &=  \frac{g^2T_F}{3N_c} \int \frac{d^3p_{{\rm cm}}}{(2\pi)^3}  \frac{d^3p_{{\rm rel}}}{(2\pi)^3} 
| \langle \psi_b | {\bf r} | \Psi_{{\bf p}_{\rm rel}} \rangle |^2
[g^{--}_E]^{>}\Big(\frac{p^2_{\rm rel}}{M}+|E_{b}|\Big) f_{Q\bar{Q}}(t, {\bf x}, {\bf p}_{{\rm cm}}, {\bf x}_{\rm rel}=0, {\bf p}_{{\rm rel}}) \nonumber \,, 
\end{align}
where $\langle \psi_b | {\bf r} | \Psi_{{\bf p}_{\rm rel}} \rangle$ is the dipole transition amplitude between a bound quarkonium state $\psi_b$ and an unbound $Q\bar{Q}$ state $\Psi_{{\bf p}_{\rm rel}}$ that is a scattering wave with momentum ${\bf p}_{\rm rel}$\@. The two-particle phase space distribution $f_{Q\bar{Q}}(t, {\bf x}, {\bf p}_{{\rm cm}}, {\bf x}_{\rm rel}=0, {\bf p}_{{\rm rel}})$ describes an unbound $Q\bar{Q}$ pair with center-of-mass (cm) position ${\bf x}_{\rm cm}={\bf x}$, momentum ${\bf p}_{{\rm cm}}$, relative position ${\bf x}_{\rm rel}=0$, and relative momentum ${\bf p}_{{\rm rel}}$\@. The relative position is fixed to be $0$ as a result of the gradient expansion used in the semiclassical limit.

\section{Chromoelectric field correlators for quarkonium in the quark-gluon plasma}

The preceding discussion shows that the dissociation/recombination dynamics of a heavy quark pair that are close together relative to the QGP temperature $T$ is determined by the correlators
\begin{align} 
\label{eq:g-definitions}
[g_{\rm adj}^{++}]^>(t) &\equiv \frac{g^2 T_F }{3 N_c}  \big\langle E_i^a(t)W^{ac}(t,+\infty) 
W^{cb}(+\infty,0) E_i^b(0) \big\rangle_T \, , \\
[g_{\rm adj}^{--}]^>(t) &\equiv \frac{g^2 T_F }{3 N_c} \big\langle W^{dc}(-i\beta - \infty, -\infty)
W^{cb}(-\infty,t)  E_i^b(t)
E_i^a(0)W^{ad}(0,-\infty)  \big\rangle_T  \, .
\end{align}
The $[g_{\rm adj}^{++}]^{>}(t)$ correlator can be interpreted as the result of calculating the medium-induced contribution to transition amplitude of a heavy quark that starts in a color singlet state and transitions to an octet state, and then taking the square of the amplitude to obtain a probability. Similarly, $[g_{\rm adj}^{--}]^{>}(t)$ describes the process of transitioning from a color octet state to a color singlet state. The labels $++/--$ describe the orientation of the Wilson lines in each correlator.

It is important to note that, while these correlation functions are defined on a thermal ensemble for the light QCD degrees of freedom (i.e., the QGP), the heavy quarks do not thermalize because $M \gg T$. However, their internal degrees of freedom, i.e., the bound/unbound states that the heavy quark pair system can form have a characteristic energy scale $|E_b| \ll M$ given by their binding energy, which can be close to the temperature of the surrounding QGP. This means that the occupancies of the different energy levels may thermalize. Therefore, it follows that these correlators provide input to the thermalization process of these internal degrees of freedom.

On a more technical level, the above discussion motivates the introduction of additional mathematical structure to make use of the thermal aspects of these correlators. In the same way as it is done for real-time correlators of fields starting from an ensemble in thermal equilibrium, one may define the Kubo-Martin-Schwinger (KMS) conjugates of $[g_{\rm adj}^{++}]^>$ by declaring $  [g_{\rm adj}^{++/--}]^<(\omega) \equiv e^{-\omega/T} [g_{\rm adj}^{++/--}]^>(\omega)$. However, because $[g_{\rm adj}^{++}]^{>}$ and $[g_{\rm adj}^{--}]^{>}(t)$ describe physical processes that are the time-reversed versions of each other, they satisfy the property~\cite{Binder:2021otw}
\begin{equation} \label{eqn:standard_kms}
[g_{\rm adj}^{++}]^>(\omega) = e^{\omega/T} [g_{\rm adj}^{--}]^>(-\omega) \,.
\end{equation}
That is to say, $[g_{\rm adj}^{++}]^{<}(\omega) \equiv [g_{\rm adj}^{--}]^>(-\omega)$ is the KMS conjugate of $[g_{\rm adj}^{++}]^{>}(\omega)$. In fact, this is the necessary KMS relation for proper thermalization of the internal degrees of freedom of the heavy quark pair (their relative motion and internal quantum numbers~\cite{Yao:2017fuc}).
It is then natural to introduce a spectral function to describe quarkonium transport:
\begin{equation} \label{eqn:rho_adj}
\rho_{\rm adj}^{++}(\omega) = [g_{\rm adj}^{++}]^>(\omega) - [g_{\rm adj}^{--}]^>(-\omega) \, .
\end{equation}
By definition, this spectral function satisfies $[g_{\rm adj}^{++}]^>(\omega) = (1 + n_B(\omega)) \rho_{\rm adj}^{++}(\omega)$, with $n_B(\omega) = (e^{\beta \omega}-1)^{-1}$ the Bose-Einstein distribution. We have kept the superscripts ``$++$'' in the label of this spectral function because one can also define $\rho_{\rm adj}^{--}(\omega) = [g_{\rm adj}^{--}]^>(\omega) - [g_{\rm adj}^{++}]^>(-\omega)$,
which contains the same information, and satisfies $\rho_{\rm adj}^{--}(\omega) = - \rho_{\rm adj}^{++}(-\omega)$ because of the time-reversal operation that relates $[g_{\rm adj}^{++}]^>$ and $[g_{\rm adj}^{--}]^>$.

\begin{figure}
    \centering
    \includegraphics[width=0.8\textwidth]{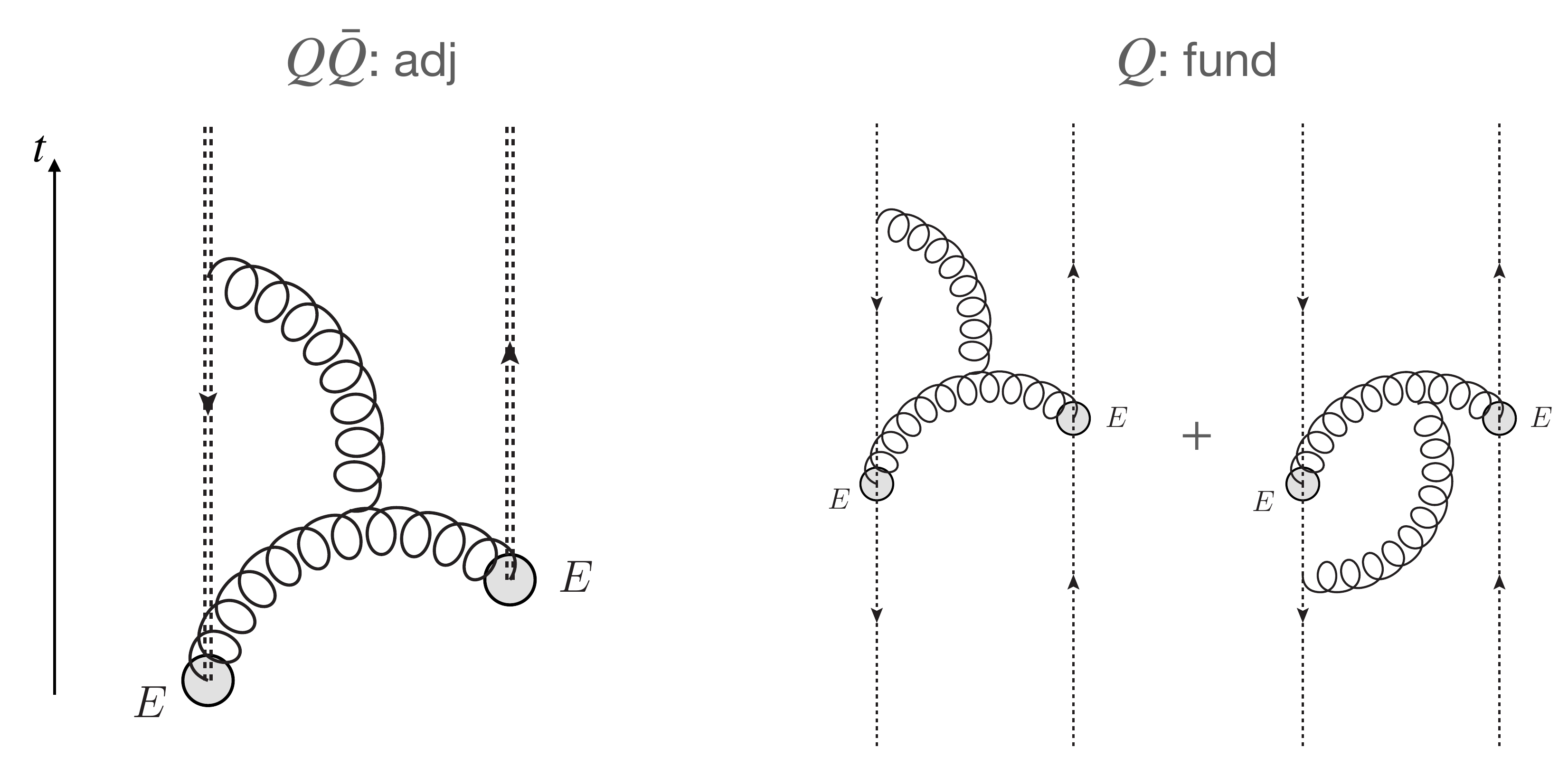}
    \caption{Diagrams that contribute to the difference between $[g_{\rm adj}^{++}]^>$ and $[g_{\rm fund}]^>$, with $[g_{\rm fund}]^{>}(\omega) = (1 + n_B(\omega)) \rho_{\rm fund}(\omega)$.}
    \label{fig:corr-diff}
\end{figure}

The fact that this spectral function is not guaranteed to be odd, since $[g_{\rm adj}^{++}]^>(\omega) \neq [g_{\rm adj}^{++}]^<(-\omega)$, highlights crucial aspects of the physical situation at hand, which are most easily illustrated by making a comparison to an analogous correlator that describes heavy quark diffusion. These differences have been emphasized in recent calculations of the quarkonium correlators~\cite{Eller:2019spw,Binder:2021otw,Nijs:2023dks}, and discussed at length in~\cite{Scheihing-Hitschfeld:2022xqx,Scheihing-Hitschfeld:2023tuz}. In perturbation theory, the difference appears at $O(g^4)$ as
\begin{align} \label{eq:rho-diff}
\Delta \rho(\omega) \equiv \big(\rho^{++}_{\rm adj}(\omega) - \rho_{\rm fund}(\omega) \big) = \frac{g^4 T_F (N_c^2-1) \pi^2}{3 (2\pi)^3}  |\omega|^3 + \mathcal{O}(g^6) \,,
\end{align}
where $\rho_{\rm fund}(\omega)$ is most easily defined as the spectral function related to the Euclidean correlator
\begin{align}
\label{eqn:Gfund}
G_{\rm fund}(\tau) = - \frac{1}{3} \frac{\big\langle {\rm Re} {\rm Tr}_c[ U(\beta,\tau) gE_i(\tau) U(\tau,0) gE_i(0) ] \big\rangle_T}{\big\langle {\rm Re} {\rm Tr}_c[U(\beta,0)] \big\rangle_T} \,,
\end{align}
which is related via the usual analytic continuation relations to its real-time counterparts~\cite{Casalderrey-Solana:2006fio}.
This difference, as Figure~\ref{fig:corr-diff} illustrates, reflects the different initial states of each physical process, as well as the different orderings of the stages of each physical process (which translates into different operator orderings in the quantum theory). Another way of stating this is that these processes have different quantum color correlation patterns.

An important outstanding problem in the study of the quarkonium correlator is to carry out a lattice QCD calculation of it. While the formulation of the lattice QCD heavy quark diffusion correlator has long been established~\cite{Caron-Huot:2009ncn}, studies about the subtleties one needs to understand to perform such a calculation for the quarkonium case are still ongoing~\cite{Scheihing-Hitschfeld:2023tuz}.

\section{A strongly coupled calculation}

At the same time as the challenges for a nonperturbative QCD determination of the correlators relevant for quarkonium are sorted out, it is also relevant to understand how the analogous correlation functions behave in similar theories where further analytical progress is feasible. Indeed, the study of heavy quark diffusion in supersymmetric $\mathcal{N}=4$ Yang-Mills theory (for brevity, $\mathcal{N}=4$ SYM)~\cite{Casalderrey-Solana:2006fio} at strong coupling in many ways preceded and paved the way for most of the more recent QCD studies, and provided a clean reference value with which to compare new results. Because of that, it is paramount to have a strongly coupled counterpart to the weakly coupled calculations that have been carried out in recent years for quarkonium transport~\cite{Brambilla:2008cx,Eller:2019spw,Binder:2021otw}.

The reason why analytic progress is possible at strong coupling in $\mathcal{N}=4$ SYM is because of the AdS/CFT correspondence, which allows one to calculate $\mathcal{N}=4$ SYM field theory observables in the limit of a large number of colors $N_c \gg 1$ and strong coupling $g^2 N_c \gg 1$ by solving classical equations of motion in a 10-dimensional gravitational theory, which is an (asymptotically) $AdS_{5} \times S_5$ spacetime.
As it turns out, the objects that i) have a simple dual gravitational description, ii) contain Wilson lines, and iii) can accommodate field strength insertions, are the (supersymmetric version of) Wilson loops~\cite{Maldacena:1997re-1998im}. Their dual description is given by extremizing the area of a surface that hangs from the boundary path determined by the Wilson loop into the bulk of $AdS_5$.

\begin{figure}
    \centering
    \includegraphics[width=0.49\textwidth]{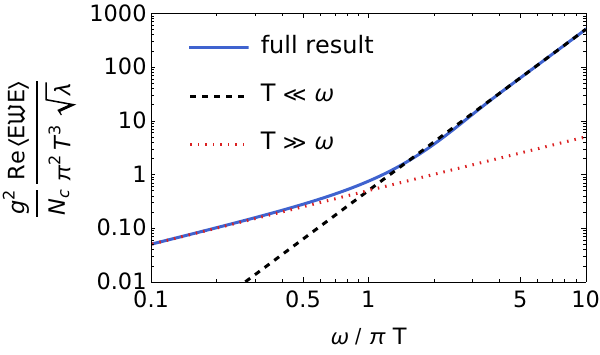}
    \includegraphics[width=0.49\textwidth]{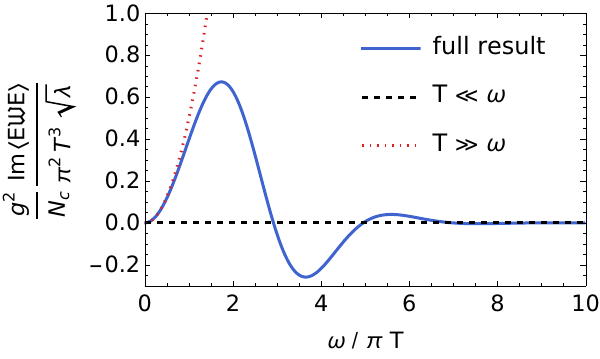}
    \caption{Real part (left) and imaginary part (right) of the time-ordered non-Abelian electric field correlator for quarkonium transport.}
    \label{fig:correlator}
\end{figure}

Following~\cite{Casalderrey-Solana:2006fio}, one can introduce electric field insertions along a Wilson loop by considering local path deformations. This amounts to considering small deformations of the dual surface, and consequently, to solving the linear response equations that determine the perturbations induced on the surface. A correlation function of two field strengths inserted along the loop is then determined by the second variational derivative of the on-shell perturbed action with respect to the boundary path. For technical details, see~\cite{Nijs:2023dks}. The result for the time-ordered correlator for quarkonium transport in $\mathcal{N}=4$ SYM, which we show in Figure~\ref{fig:correlator}, is given by
\begin{align}
     \frac{g^2}{N_c} \int_{-\infty}^\infty dt \, e^{i\omega t}  \langle \hat{\mathcal{T}} E_i^a(t) \mathcal{W}^{ab}_{[t,0]} E_j^b(0) \rangle_T =  \delta_{ij}  \frac{ (\pi T)^3 \sqrt{\lambda}}{{ 4}\pi} \left( \frac{-i}{F^-_{|\omega|}(0)} \frac{\partial^3 F^-_{|\omega|}}{\partial \xi^3}(0) \right) \, ,
\end{align}
where $F^-_\omega$ is the regular solution of the differential equation
\begin{align} \label{eq:F-thermal}
    \frac{\partial^2 F^-_\omega}{\partial \xi^2} - 2 \left[ \frac{1 + \xi^4}{\xi(1-\xi^4)} - \frac{i \Omega \xi^3}{1-\xi^4} \right] \frac{\partial F^-_\omega}{\partial \xi} + \left[ \frac{i \Omega \xi^2}{1-\xi^4} + \frac{\Omega^2 (1 - \xi^6) }{(1-\xi^4)^2} \right] F^-_\omega = 0 \, ,
\end{align}
with $\Omega = \omega/(\pi T)$. This allows one to determine the spectral function for quarkonium transport $\rho_{\rm adj}^{++}(\omega)$. The main result of this calculation is that the quarkonium transport coefficients in the Lindblad equation vanish $\kappa_{\rm adj}^{\mathcal{N}=4 } = \gamma_{\rm adj}^{\mathcal{N}=4 } = 0 $ in $\mathcal{N}=4$ SYM at strong coupling in the large $N_c$ limit. One can verify that the dynamics in the semiclassical rate Eq.~\eqref{eqn:rate} are also trivial in this limit.

%, which is apparent in Figure~\ref{fig:correlator}. For the proof of this see~\cite{Nijs:2023dks}.

\section{Conclusions}

We have discussed how the correlator of two chromoelectric fields connected by an adjoint Wilson line enters the dynamics of quarkonium in a thermal medium, and shown the calculation results of the analogous correlation function in $\mathcal{N}=4$ SYM at strong coupling. Given past successes of holographic calculations describing the quark-gluon plasma, the next step is to study the phenomenology of quarkonium dissociation and recombination using this correlation function. However, because the transport coefficients vanish in the descriptions considered herein, to leading order the dynamics induced by this correlator are trivial. Therefore, further theoretical developments will be needed to assess the effects of a strongly coupled plasma on quarkonium transport.

\acknowledgments

This work is supported by the U.S.~Department of Energy, Office of Science, Office of Nuclear Physics under grant Contract Number DE-SC0011090\@. XY is supported by the U.S. Department of Energy, Office of Science, Office of Nuclear Physics, InQubator for Quantum Simulation (IQuS) (https://iqus.uw.edu) under Award Number DOE (NP) Award DE-SC0020970 via the program on Quantum Horizons: QIS Research and Innovation for Nuclear Science.

%\begin{align} \label{eq:W-loop-S}
% W_{\rm BPS}[\mathcal{C};\hat{n}] = \frac{1}{N_c} {\rm Tr}_{\rm color} \! \left[ \mathcal{P} \exp \left( ig \oint_{\mathcal{C}} ds \, T^a \left[  A^a_\mu \, \dot{x}^\mu + \hat{n}(s) \cdot {\vec{\phi}}^a \sqrt{\dot{x}^2} \right] \right) \right] \, ,
% \end{align}
% where $\dot{x}^\mu(s) = dx^\mu(s)/ds$ and $\vec{\phi} = (\phi^1, \ldots, \phi^6)$ are the six Lorentz scalar fields in the adjoint representation of SU($N_c$) intrinsic to $\mathcal{N}=4$ SYM. These scalars enter the Wilson loop coupled to a direction $\hat{n}(s) \in S_5$ that specifies the direction along which the string ``pulls'' the heavy quark.

\end{document}